\documentclass{llncs}
\usepackage{graphicx}
\usepackage{amsmath}
\usepackage{amssymb}
\usepackage{hyperref}
\hypersetup{
    colorlinks=true,
    linkcolor=blue,
    citecolor=black,
    urlcolor=blue,
}
\usepackage{CJKutf8}
\usepackage[misc]{ifsym}
\usepackage[T1]{fontenc}
\begin{document}\sloppy
\title{PatientEG Dataset: Bringing  Event Graph Model with Temporal Relations to Electronic Medical Records}
%
%

\author{Xuli Liu\inst{1} 
\and Jihao Jin\inst{1}
\and Qi Wang \inst{1}
\and Tong Ruan \inst{1}
\and Yangming Zhou \inst{1}
\and Daqi Gao \inst{1}$^{\textrm{(\Letter)}}$
\and Yichao Yin \inst{2}
}
\authorrunning{Liu. et al.}
%
\institute{School of Information Science and Engineering, East China University of Science and Technology, Shanghai 200237, China\\
\email{\{615877848,654091337\}@qq.com, dsx4602@163.com, \{ruantong,ymzhou,gaodaqi\}@ecust.edu.cn}
\and Shanghai Shuguang Hospital, Shanghai 200021, China\\
\email{sandos\_os\_2002@163.com}
}
\let\oldmaketitle\maketitle
\renewcommand{\maketitle}{\oldmaketitle\setcounter{footnote}{0}}
\maketitle  

\begin{abstract}
Medical activities, such as diagnoses, medicine treatments, and laboratory tests, as well as temporal relations between these activities are the basic concepts in clinical research. However, existing relational data model on electronic medical records (EMRs) lacks explicit and accurate semantic definitions of these concepts. It leads to the inconvenience of query construction and the inefficiency of query execution where multi-table join queries are frequently required. In this paper, we propose a patient event graph (PatientEG) model to capture the characteristics of EMRs. We respectively define five types of medical entities, five types of medical events and five types of temporal relations. Based on the proposed model, we also construct a PatientEG dataset with 191,294 events, 3,429 distinct entities, and 545,993 temporal relations using EMRs from Shanghai Shuguang hospital. To help to normalize entity values which contain synonyms, hyponymies, and abbreviations, we link them with the Chinese biomedical knowledge graph. With the help of PatientEG dataset, we are able to conveniently perform complex queries for clinical research such as auxiliary diagnosis and therapeutic effectiveness analysis. In addition, we provide a SPARQL endpoint to access PatientEG dataset and the dataset is also publicly available online. Also, we list several illustrative SPARQL queries on our website.
\end{abstract}

\textbf{Resource type:} Dataset

\textbf{Permanent URL:} \url{http://peg.ecustnlplab.com}

\section{Introduction}
The quantity of electronic medical records (EMRs) has an explosion since the hospital information systems have been widely adopted a decade ago. EMRs are the main carrier for recording various medical activities on patients in hospital, such as diagnoses, medicine treatments, and laboratory tests. Medical activities and temporal relations in EMRs are fundamental concepts for clinical informatics applications such as auxiliary diagnosis \cite{ben2015improving}, therapeutic effectiveness analysis \cite{loughlin2018effectiveness} and mortality risk prediction \cite{ghassemi2014unfolding}. 

However, most of EMRs are stored in relational databases, which are not well designed for clinical research: 1) The relational data model lacks of an explicit and accurate semantic definition for medical activities. 2) Many clinical tasks involve various medical activities distributed among multiple tables, and it is inefficient for query executions where frequent queries of multi-table join are involved. 3) The temporal relations between medical activities are important in EMRs, and can be employed to track the health status of the patient to find the cause of the disease and analyze the effectiveness and side effects of the treatment. But the relational data model cannnot express temporal relations explicitly.

Compared to the existing relational data model, graph representation for EMRs is a better choice. Graph 
representation has essential advantages of intuitively representing massive relations between heterogeneous things. It is closer to human cognition, making it convenient to construct and efficient to execute complex queries. Moreover, graph representation has a flexible extension to support various domain ontology. Therefore, a graph representation composed of medical activities and temporal relations is urgently desired for clinical informatics applications. Recently, event-centric graph representation, which has benefits in representing event and temporal information, attaches importance to many fields, such as social network recommendation \cite{pham2015general}, news event extraction \cite{rospocher2016building}, and scientific event research \cite{fathalla2018eventskg}. Furthermore, Hage et al. \cite{van2011design} created an ontology to model events in various subject domains. Gottschalk and Demidova \cite{gottschalk2018eventkg} built a multilingual event knowledge graph from Wikidata \cite{erxleben2014introducing}, DBpedia  \cite{lehmann2015dbpedia}, and YAGO \cite{mahdisoltani2013yago3}. Inspired by their success, we attempt to apply event-centric graph representation for EMRs. Concretely, we propose a patient data representation based on event graph model, and respectively define five types of medical entities, five types of medical events, as well as five types of temporal relations for EMRs on this representation. 

Medical entities derived from EMRs are isolated and  their values are none normalized. Synonymies, hyponymies and abbreviations frequently occur in medical entities. Query construction becomes complex since a synset instead of a word should be used in the query. Furthermore, a typical query in clinical research may query on a sort of diseases or medicines, such as, ``western medicine'' or ``ACEI medicine''. In both cases, a terminology graph and linkage with the graph are required. In recent years, many linked open biomedical knowledge graphs are published using Resource Description Framework (RDF) \cite{world2014rdf} format. Godoy et al. \cite{garcia2013sharing} provide the largest network of Linked Data for the Life Sciences. We have released a Chinese biomedical knowledge graph (CBioMedKG) in our prior work \cite{ruan2017automatic}. Also we extracted 26,821 symptoms, 292 departments, 32,956 diseases, 67,712 medicines, and 7704 assays as well as more than 20 categories of relations between symptoms and the above related entities from mainstream healthcare websites and Chinese encyclopedia sites. For medical entities in EMRs, it is necessary to link them with CBioMedKG in order to utilize many synonyms, hyponymies and abbreviations of entities.

In this paper, we propose a patient event graph  (PatientEG) representation upon Simple Event Model (SEM) \cite{van2011design} to model medical activities and temporal information for EMRs. Based on the proposed model, a dataset is constructed from EMRs to facilitate clinical research. In addition, we link entities in the dataset with CBioMedKG to normalize entity values and provide more medical information. The contributions of this paper are as follows:

\begin{itemize}
  \item We propose a PatientEG model which can be leveraged on EMRs to carry out clinical research conveniently.
  \item We construct a PatientEG dataset from EMRs based on the proposed model, and link entities with CBioMedKG to construct queries on patients with domain knowledge.
  \item We publish PatientEG dataset as linked data on Web and provide an online access via SPARQL endpoint. We also list several illustrative SPARQL queries on our website.
\end{itemize}

\section{Patient Event Graph Representation} \label{model} 
To capture the characteristics of EMRs, a new representation is proposed to model  medical activities and temporal relations between activities. In this section, we first give the definition of our PatientEG model, then we introduce medical entities, medical events, temporal relations between events, and a constraint of temporal relations in details.

\begin{figure}
\vspace{-0.4cm}
\setlength{\belowcaptionskip}{-0.2cm}
\includegraphics[width=\textwidth]{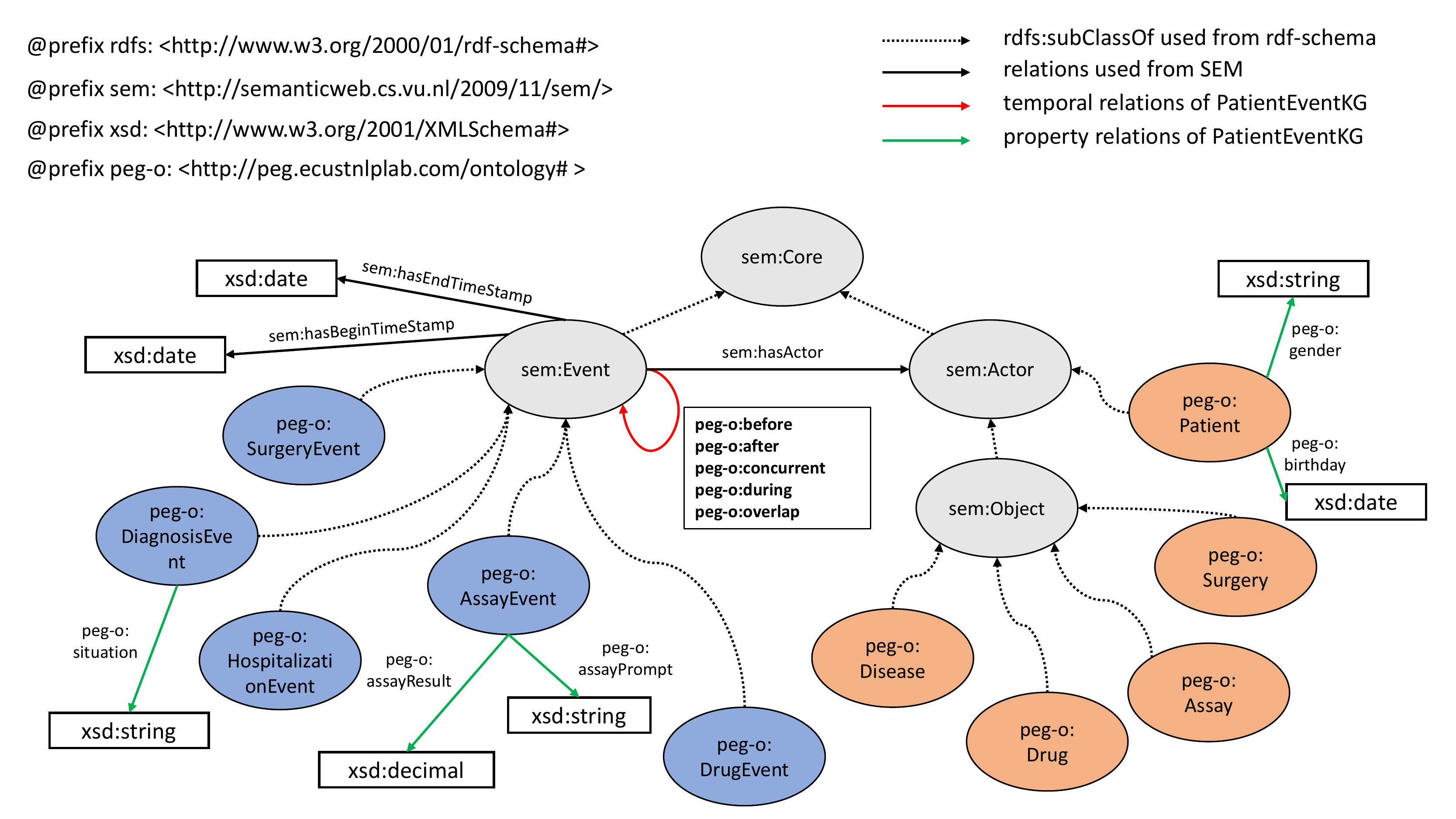}
\caption{The PatientEG schema based on SEM. Grey ellipses denote classes used from SEM. Blue and orange ellipses respectively represent the classes of medical events and entities. Arrows with dashed line represent rdfs:subClassOf. Green arrows denote property relations . Red arrow indicates five temporal relations between events.} \label{fig2_1}
\end{figure}

\paragraph{\textbf{PatientEG model:}}
The schema we design for PatientEG is based on Simple Event Model (SEM), a domain-independent event model which provides a flexible framework for building generic event-centric datasets.  SEM uses several core classes, types and constrains to describe events. The class \texttt {sem:Event}\footnote{sem is the prefix within SEM namespace, \url{https://semanticweb.cs.vu.nl/2009/11/sem/}} indicates what happened, the class \texttt{sem:Actor} represents who or what participated, and the class \texttt{sem:Time} represents when an event took place. To cover the information of medical activities and temporal relations from EMRs, we further add several necessary properties, classes, and relations. The schema of PatientEG is shown as Fig.~\ref{fig2_1}

\paragraph{\textbf{Medical entities:}} Medical entities in EMRs mainly include patients, diseases, drugs, assays and surgeries. In SEM, the class \texttt{sem:Actor} holds entities that take part in an event, and the class \texttt{sem:Object} represents passive and inanimate \texttt{sem:Actor}. In our PatientEG model,  \texttt{peg-o:Patient}\footnote{peg-o is the prefix of  ontology vocabulary identifier within PatientEG namespace, \url{http://peg.ecustnlplab.com/ontology\#}} representing patient entities, is defined as the subclass of \texttt{sem:Actor}. \texttt{peg-o:Disease}, \texttt{peg-o:Drug}, \texttt{peg-o:Assay}, and \texttt{peg-o:Surgery} are defined as the subclasses of \texttt{sem:Object} to indicate the entities of diseases, drugs, assays and surgeries. All patient entities are uniquely identified as resources within the namespace \texttt{peg-r\footnote{peg-r is the prefix of resource identifier within PatientEG namespace, \url{http://peg.ecustnlplab.com/resource/}}}. The properties of patient entities contain gender and birthday. We respectively use \texttt{peg-o:gender} and \texttt{peg-o:birthday} to indicate gender and birthday properties. 
\begin{figure}
\vspace{-0.4cm}
\setlength{\belowcaptionskip}{-0.2cm}
\begin{center}
\includegraphics[width=0.8\textwidth]{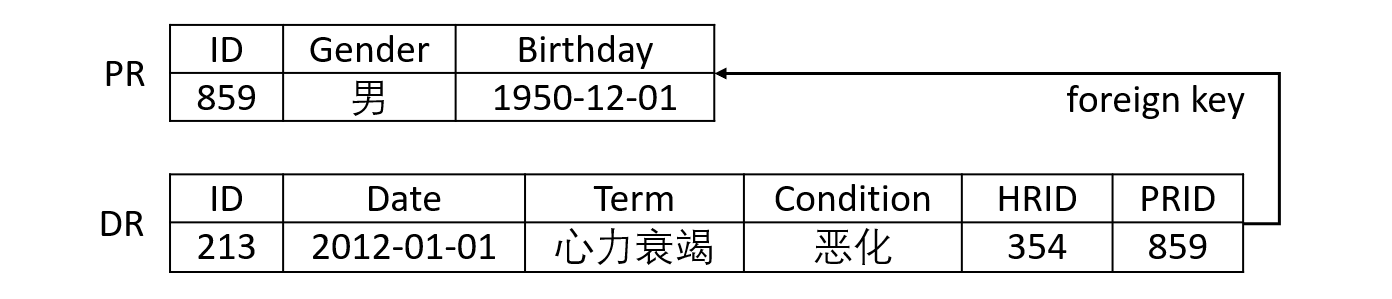}
\begin{CJK*}{UTF8}{gbsn}
\caption{An example of diagnosis activities: The male (``男'') patient numbered 859 was diagnosed with heart failure (``心力衰竭'') on January 1, 2012, and his condition deteriorated (``恶化'') at the time of diagnosis.} 
\end{CJK*}
\label{fig2_2}
\end{center}
\end{figure}
\paragraph{\textbf{Medical events:}} EMRs contain a large number of medical activities that arerecorded in various tables of relational databases. These activities are the basis for our establishment of medical events. A real-world example of diagnosis activities implicitly appearing in EMRs is shown as Fig.~\ref{fig2_2}, which involved two records, i.e. patient record (PR) and diagnosis record (DR).

\begin{CJK*}{UTF8}{gbsn}
\begin{figure}
\vspace{-0.4cm}
\setlength{\belowcaptionskip}{-0.4cm}
\begin{center}
\includegraphics[width=0.8\textwidth]{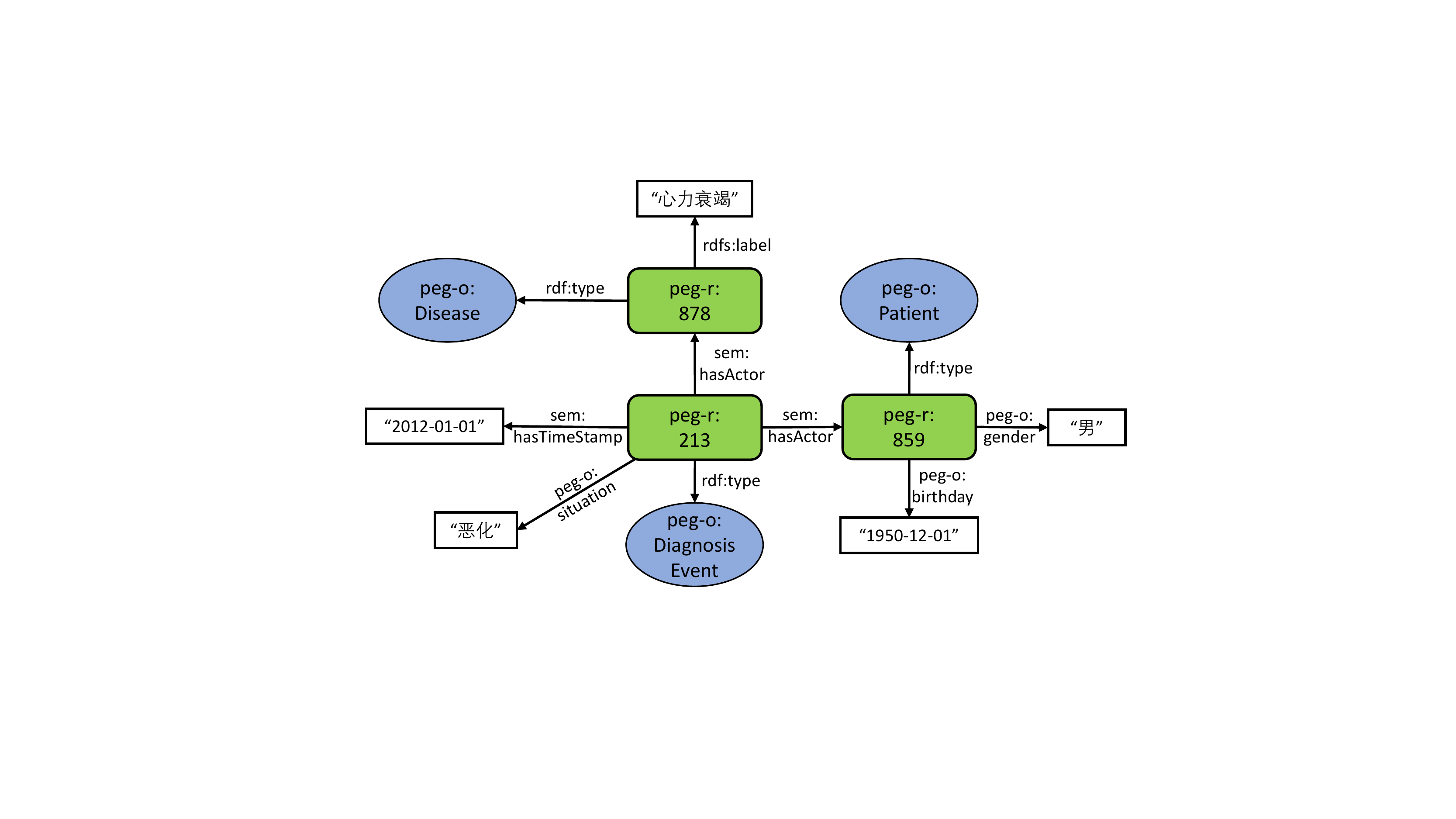}
\caption{A diagnosis event graph. It consists of three parts presented with green rounded rectangles, blue ellipses and white rectangles. The green rounded rectangle \texttt{peg-r:213} denotes the diagnosis event resource, \texttt{peg-r:859} denotes the male (``男'') patient resource, and \texttt{peg-r:878} denotes the disease entity resource of heart failure (``心力衰竭'').} 
\label{fig2_3}
\end{center}
\end{figure}
\end{CJK*}
SEM provides a skeleton for modeling event. \texttt{sem:Event} representing things that happen, and is the central class of SEM. We define five subclasses of \texttt{sem:Event}, i.e.  \texttt{peg-o:HospitalizationEvent},  \texttt{peg-o:DiagnosisEvent}, \texttt{peg-o:DrugEvent}, \texttt{peg-o:AssayEvent}, and \texttt{peg-o:SurgeryEvent} to denote various medical events. As in Fig.~\ref{fig2_3}, the diagnosis event \texttt{peg-r:213} is an instance of \texttt{peg-o:DiagnosisEvent} and we use \texttt{rdf:type} to denote this relation. Two actors are involved in the diagnosis event, \texttt{peg-r:859} denotes a male patient numbered 859 and \texttt{peg-r:878} is a disease entity named ``heart failure''. The type of the former entity is \texttt{peg-o:Patient}, and the latter is \texttt{peg-o:Disease}. SEM employs \texttt{sem:hasActor} to establish connection between \texttt{sem:Event} and \texttt{sem:Actor}. Thus, we can obtain two triples, \\ \centerline{$<$ \texttt{peg-o:DiagnosisEvent}, \texttt{sem:hasActor}, \texttt{peg-o:Patient}$>$} \\and\\ \centerline{$<$ \texttt{peg-o:DiagnosisEvent}, \texttt{sem:hasActor}, \texttt{peg-o:Disease}$>$,}\\ as part of the ontology of PatientEG.

All medical events in EMRs have accurate timestamps, we no longer use the class \texttt{sem:Time} to represent occurrence time of events, but add  properties to \texttt{sem:Event} directly. We use \texttt{sem:hasBeginTimeStamp} to indicate start time of events, and \texttt{sem:hasEndTimeStamp} to indicate end time of events, as shown in Fig.~\ref{fig2_1}. In addition, medical events contain other properties. For example, diagnosis events include the conditions of illness at the time of diagnosis, assay events include assay results and prompts. In PatientEG, we use \texttt{peg-o:situation} to indicate the conditions of illness, \texttt{peg-o:assayResult} to indicate assay results, and \texttt{peg-o:assayPrompt} to indicate assay prompts. Fig.~\ref{fig2_3} illustrates a graph of a diagnosis event.

\begin{figure}
\vspace{-0.4cm}
\setlength{\belowcaptionskip}{-0.4cm}
\begin{center}
\includegraphics[width=0.8\textwidth]{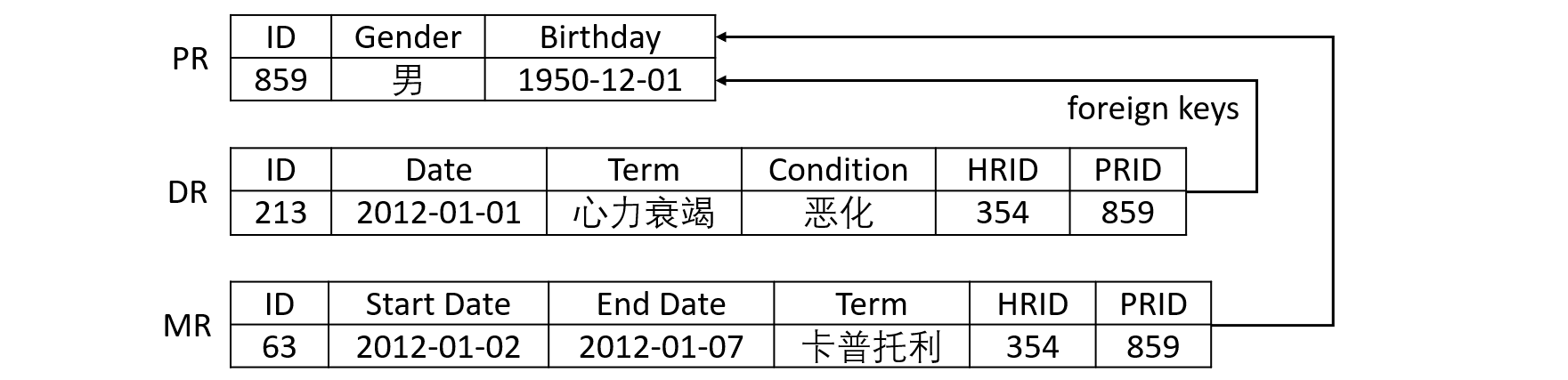}
\caption{An example of underlying temporal relations between records.: The patient numbered 856 was diagnosed with heart failure (``心力衰竭'') on January 01, 2012, and was treated with captopril (``卡普托利'') from January 02, 2012 to January 07, 2012.}
\label{fig2_4}
\end{center}
\end{figure}

\paragraph{\textbf{Temporal relations between medical events:}}
The fields in various records, such as ``Date'', ``StartDate'' and ``EndDate'', represent the occurrence time of medical activities, also implicitly indicate the temporal information between different activities . An example of underlying temporal information in EMRs is shown as Fig.~\ref{fig2_4}, which involves three records: patient record (PR), diagnosis record (DR), and medication record (MR).
\begin{CJK*}{UTF8}{gbsn}
\end{CJK*}

\begin{figure}
\vspace{-0.4cm}
\setlength{\belowcaptionskip}{-0.4cm}
\begin{center}
\includegraphics[width=\textwidth]{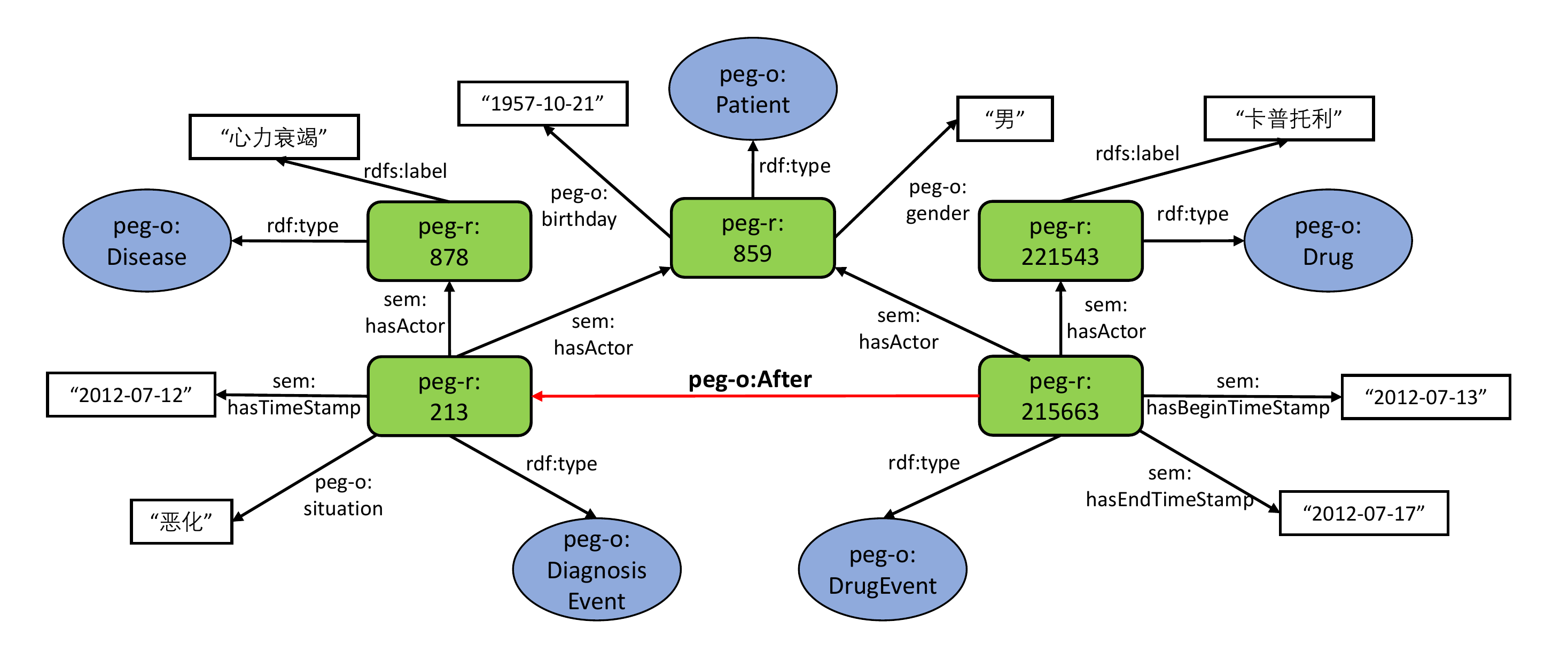}
\caption{A graph of \texttt{peg-o:After} temporal relation. Red arrow denotes the \texttt{After} relation between the drug event \texttt{peg-r:215663} and the diagnosis event \texttt{peg-r:213}.} \label{fig2_5}
\end{center}
\end{figure}

This example is composed of two medical events, 1) A diagnosis event where a patient numbered 859 was diagnosed with heart failure on January 01, 2012. 2) A drug event where the patient took captopril for treatment from January 02, 2012 to January 07, 2012. The diagnosis event occurred before the drug event. Conversely, the drug event occurred after the diagnosis event. We employ \texttt{peg-o:Before} and \texttt{peg-o:After} to denote the relations of \texttt{Before} and \texttt{After}. In addition to above two temporal relations, there are three meaningful temporal relations:
\begin{itemize}
    \item  \texttt{Concurrent} relation means that two medical events occur at the same time. For example, it occurs in EMRs that ``\textit{a patient numbered 583 was injected 'Adrenaline' and examined 'Blood pressure' on October 10, 2012.}'' The occurrence time of the drug event where the patient was injected 'Adrenaline' is equal to the assay event where the patient examined 'Blood pressure'.
    \item \texttt{During} relation means that the medical event \textit{eventA} occurs during the ongoing medical event \textit{eventB}, and the end occurrence time of \textit{eventA} is earlier than the end occurrence time of \textit{eventB}.
    \item \texttt{Overlap} relation is similar to the \texttt{During} relation except that the end occurrence time of \textit{eventA} is later than the end occurrence time of \textit{eventB}.
\end{itemize}
We adopt additional temporal relations \texttt{peg-o:Concurrent}, \texttt{peg-o:During}, and \texttt{peg-o:Overlap} within the namespace of PatientEG. Fig.~\ref{fig2_5} shows a graph of temporal relations between medical events.
\paragraph{\textbf{Constraint of temporal relations:}}
There is a constraint with temporal relations that only the medical events occurring on the same patient are time-comparable. In our PatientEG, temporal relations between medical events are established with the same patient.

\section{PatientEG Construction}\label{construction}  
\begin{figure}
\includegraphics[width=\textwidth]{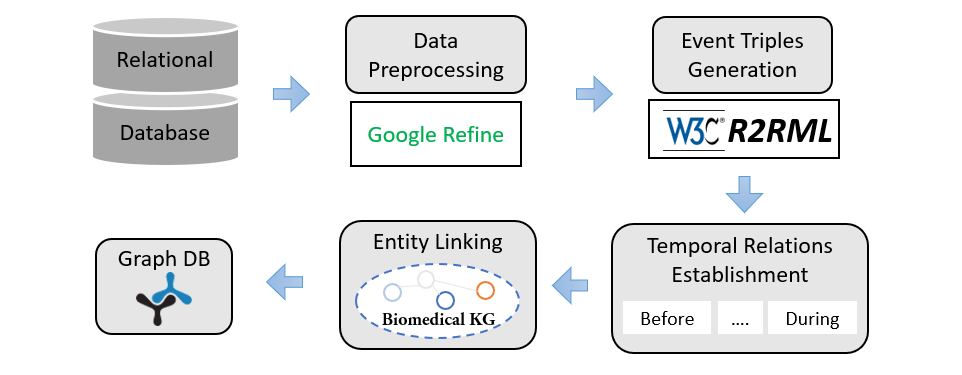}
\caption{The workflow of construction for PatientEG dataset. It is composed of four steps: data preprocessing, event triples generation, temporal relation establishment, and instance matching.} \label{fig3_1}
\end{figure}

In this section, we describe a workflow of constructing PatientEG dataset. As shown in Fig.~\ref{fig3_1}, the workflow consists of four steps, namely data preprocessing, event triples generation, temporal relation establishment, and instance matching. Data preprocessing step aims to address messy data such as null values and inconsistency of units. Event triples generation step offers a normative process mapping relational data to RDF tripes. Temporal relation establishment step builds five types of temporal relations among medical events. We bridge entities in PatientEG dataset and CBioMedKG in instance matching step. Next, We will elaborate on the four steps in following subsections.
\subsection{Data preprocessing}
Due to the defects of data management in hospital, several data quality issues, such as missing values, non-uniform units, and non-standard values, exist in EMRs. For example, assay result includes two qualitative values, positive and negative. However, there are many different expressions, such as ``negative (-)'', ``positive (+)'', ``(+)'', and ``(-)'', which require converting to standard values. In addition, the unit of the same assay result values may be inconsistent. For instance, the ``hemoglobin'' result values use the units of g/dL and g/L. We need standardize them to a same unit, and the result values should also be converted according to conversion relations between their units. Another example occurs in diagnosis records: If a patient was diagnosed with multiple diseases, the names of diseases are recorded in a same cell split by commas, spaces, and semicolons. We need disassemble them such that each line in records stores only one disease. In this paper, Google Refine\footnote{\url{http://openrefine.org/}} assists us to process EMRs with a variety of ways, such as converts, splits and merges.

\subsection{Event triples generation}
\begin{figure}
\includegraphics[width=\textwidth]{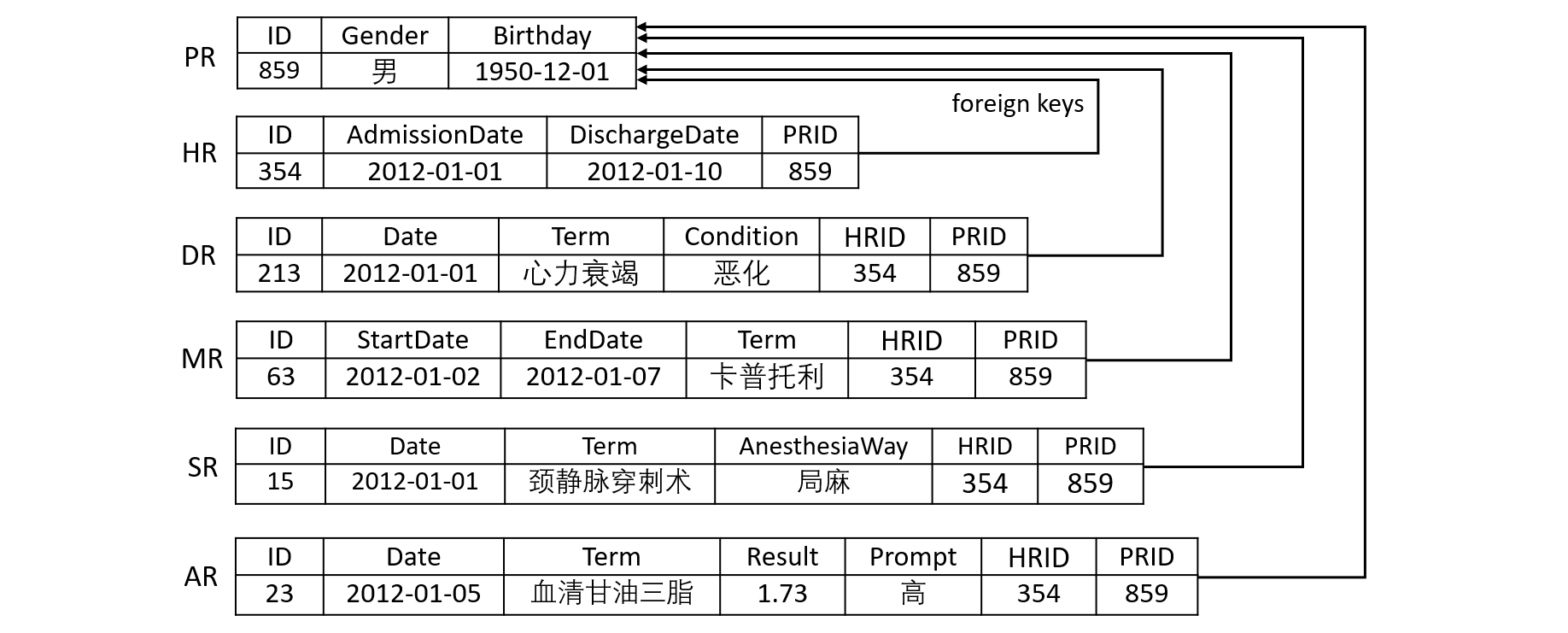}
\caption{An excerpt of EMRs in a relational database.} \label{fig3_2}
\end{figure}

To achieve availability and reusability, we use the RDF standard to generate patient and event triples from EMRs. The vast majority of EMRs are stored in relational databases. Foreign keys in relational databases establish references from rows in a table to exactly rows in another table. An excerpt of EMRs in a relational database is shown as Fig.~\ref{fig3_2}, which includes a patient record (PR), a hospitalization record (HR), a diagnosis record (DR), a medication record (MR), an assay record (AR), and a surgery record (SR).

To convert relational data into RDF triples, we empoly the W3C recommended RDB2RDF mapping standard \cite{arenas2012direct} to directly map relational data to RDF triples. The direct mapping standard defines an RDF graph representation of the data in a relational database, which takes as input a relational database (data and schema), and generates an RDF graph that is called direct graph. The direct mapping is described as follows:
\begin{itemize}
  \item Each logical table is mapped to RDF using a triples map.
  \item Triples map is a rule that maps each row in the logical table to a number of RDF triples. Each triples map includes a subject map and multiple predicate-object maps. 
  \item The subject map generates the subject of all RDF triples that will be generated from a logical table row. And the subjects are IRIs that are generated from the primary key column of the table. 
  \item The multiple predicate-object maps that in turn consist of predicate maps and object maps (or referencing object maps).
  \item Triples are produced by combining the subject map with a predicate map and object map, and applying these three to each logical table row.
\end{itemize} 
We provide five triples maps corresponding to PRs, HRs, DRs, ERs, and SRs to generate event triples on our website.

\subsection{Construction of temporal relations}
In section \ref{model}, we have defined five temporal relations,  namely \texttt{Before}, \texttt{After}, \texttt{Concurrent}, \texttt{During}, and \texttt{Overlap}. We present the process of building temporal relations in detail in this subsection. 

The occurrence time of all medical events can be divided into two types: time point and time period. For example, a diagnosis event occurs at a certain time, the type of occurrence time is time point. While a drug event lasts for a period of time and the type of occurrence time is time period. Considering the complication of constructing five temporal relations together, we process them step by step:
\begin{itemize}
    \item First, we build the temporal relations of \texttt{Before}, \texttt{After}, and \texttt{Concurrent} between events whose occurrence time type is time point.
    \item Second, we build \texttt{Before}, \texttt{After}, and \texttt{Concurrent} relations  between events whose occurrence time type is time period.
    \item Then, we construct the relations of \texttt{Before}, \texttt{After}, and \texttt{During} between events whose occurrence time type is time point and events whose occurrence time type is time period.
    \item Finally, we build the relations of \texttt{During} and \texttt{overlap} between events whose occurrence time type is time period.
\end{itemize}
Due to the fact that it is redundant to construct all the relations among medical events, we only construct some of them, and the others can be inferred. An example of the inference rule is shown as follows. Given three medical events named \textit{eventA}, \textit{eventB}, and \textit{eventC}. According to occurrence time, if \textit{timeA} $>$ \textit{timeB} $>$ \textit{timeC}, we only need to build \texttt{Before} and \texttt{After} relations between \textit{eventA} and \textit{eventB},  as well as \textit{eventB} and \textit{eventC}. The relations of \texttt{Before} and \texttt{After} between \textit{eventA} and \textit{eventC} can be inferred through \textit{eventB}, and we no longer build direct relations.

\subsection{Instance matching} 
\begin{figure}
\includegraphics[width=\textwidth]{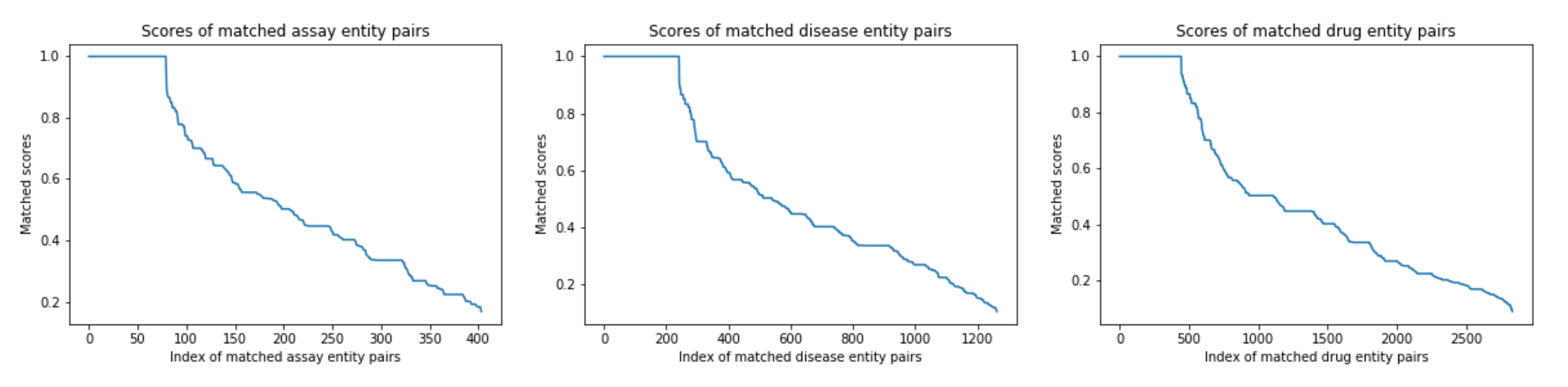}
\caption{Matchability scores of entities. Each subfigure corresponding to a type of entities. The horizontal axis represents the index of matched entity pairs. The vertical axis represents the matchability scores.} \label{fig3_4}
\end{figure}

Let $m$ be the entity in PatientEG dataset, $e$ be the entity in CBioMedKG, the matchability $score(m,e)$ between them is computed as:
\begin{equation}
score(m,e) = average(levenshtein(m,e) + jaccard(m,e) + lcs(m,e))
\end{equation}
where $levenshtein$, $jaccard$ and $lcs$ denote the Levenshtein Similarity, Jaccard Similarity and Longest Common Subsequence Similarity, respectively.
For each entity $m_i$ in PatientEG dataset, we select the entity $e_j$ with the same type as $m_i$ in CBioMedKG which has the highest matchability $score(m_i,e_j)$ as the candidate aligned term. The matchability scores of each type of entities are shown in Fig. ~\ref{fig3_4}.

\begin{figure}
\includegraphics[width=\textwidth]{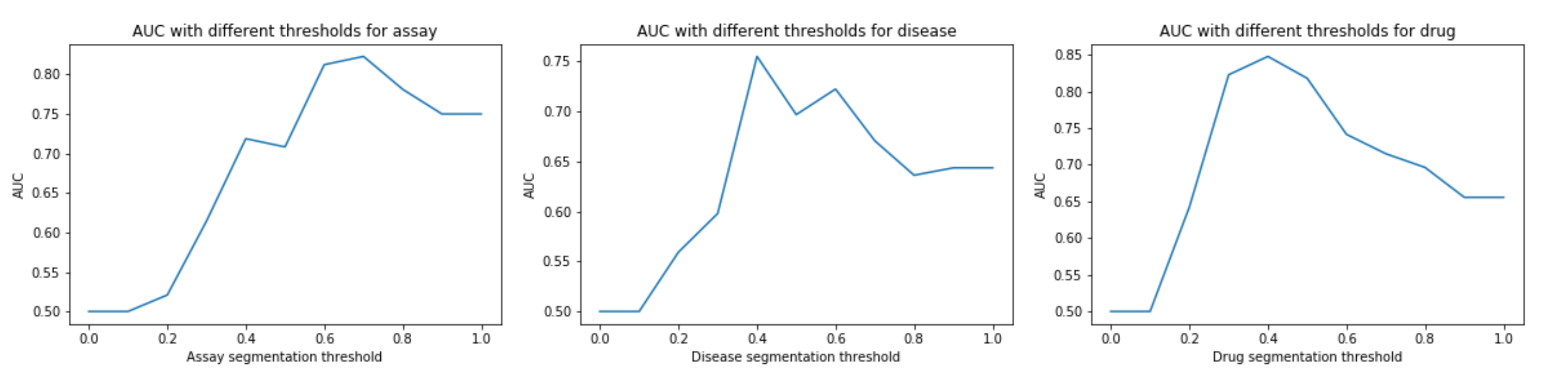}
\caption{AUC with different thresholds. Each subfigure corresponding to a type of entities. The horizontal axis represents the threshold. The vertical axis represents the AUC value.}\label{fig3_5}
\end{figure}

The candidate aligned terms obtained may be incorrect, so a threshold for each type of entities is required. Only the entity pairs with matchability above the threshold are linked. Each threshold is determined on a small-scale verification set based on the uniform sampling of matchability scores. The sampling ratio is 10\%, and the candidate aligned term is marked by 6 experts. We use AUC\footnote{Area under the Curve of ROC} to measure the effect of the different thresholds, as shown in Fig. \ref{fig3_5}. We apply the optimal thresholds to all entities, and the final link results are shown in Table.~ \ref{tab1}.

\begin{table}[]
\caption{Linking result with the entities of diseases, drugs and assays from PatientEG dataset to CBioMedKG.}\label{tab1}
\centering
\begin{tabular}{|l|c|c|c|c|c|c|}
\hline
Entity type & Threshold & AUC & Link rate \\
\hline
Disease & 0.40 & 0.76 & 0.801 \\
\hline
Drug & 0.40 & 0.85 & 0.799 \\
\hline
Assay & 0.70 & 0.82 & 0.642 \\
\hline
\end{tabular}
\end{table}

\section{PatientEG dataset characteristics} \label{statisticsAndcharacteristics}
\paragraph{\textbf{Data size:}} We use part of EMRs in Shanghai Shuguang Hospital, which include 304 patient records, 1,433 hospitalization records, 7,319 diagnosis records, 125,757 assay records, 11 surgery records, and 56,774 drug records.
A total of 1,144,754 triples were extracted from EMRs, which include 7,319 diagnosis events, 125,757 assay events, 56,774 drug events, and 11 surgery events. We bulid 142,480 \texttt{Before} relations, 142,480 \texttt{After} relations, 180,517 \texttt{Concurrent} relations, 69,236 \texttt{During} relations, and 10,841 \texttt{Overlap} relations. For entities in PatientEG dataset, 5,866 diseases, 100,456 assays, and 36,477 drugs are linked to the CBioMedKG.

\paragraph{\textbf{Availability:}}
We released PatientEG ontology and dataset under a \textit{Creative Commons Attribution 4.0} license and they can be downloaded from our website. We also provide a helpful instruction about PatientEG. Online query and several illustrative queries are available through a SPARQL endpoint on our website.
\paragraph{\textbf{Reusability:}} The reusability of PatientEG can be seen from two perspectives. First, the ontology of PatientEG is built upon a general event model, which provides a flexible expansion for other biomedical knowledge. For example, we can easily complement the causal relationship between diseases and drugs. Second, PatientEG dataset follows best practices in data publishing and adopts a popular open standard, which makes it simple to fuse other knowledge bases.

\section{Example applications on PatientEG dataset}
To intuitively present the practical applications of our PatientEG dataset, we collect a series of scenarios in clinical applications from doctors. According to the complexity of the queries, we divide these queries into three categories: single event (SE), multiple events (ME) and constrained multiple events (CME). SE means that a scenarios only contains a simple event. ME means that a scenarios composed of several events, but dose not contain temporal relations. CME means that a scenarios not only includes various events but also contains complex temporal relations.
\begin{table}[]
\caption{Examples of clinical problems on PatientEG dataset}\label{tab2}
\centering
\begin{tabular}{|p{2cm}|p{10cm}|}
\hline
Category & Clinical problems \\
\hline
SE & \textit{List patients diagnosed with coronary heart disease.} \\
\hline
ME & \textit{Which female patients were diagnosed cancer of the stomach and were treated with ftorafur and uramustine tablets?} \\
\hline
CME & \textit{How many male patients were diagnosed with coronary heart disease during hospitalization and then took captopril for treatment, during the medical  treatment, their globulin returned normal?}\\
\hline
\end{tabular}
\end{table}
\begin{CJK*}{UTF8}{gbsn}
\begin{figure}
\vspace{-0.4cm}
\setlength{\belowcaptionskip}{-0.6cm}
\includegraphics[width=\textwidth]{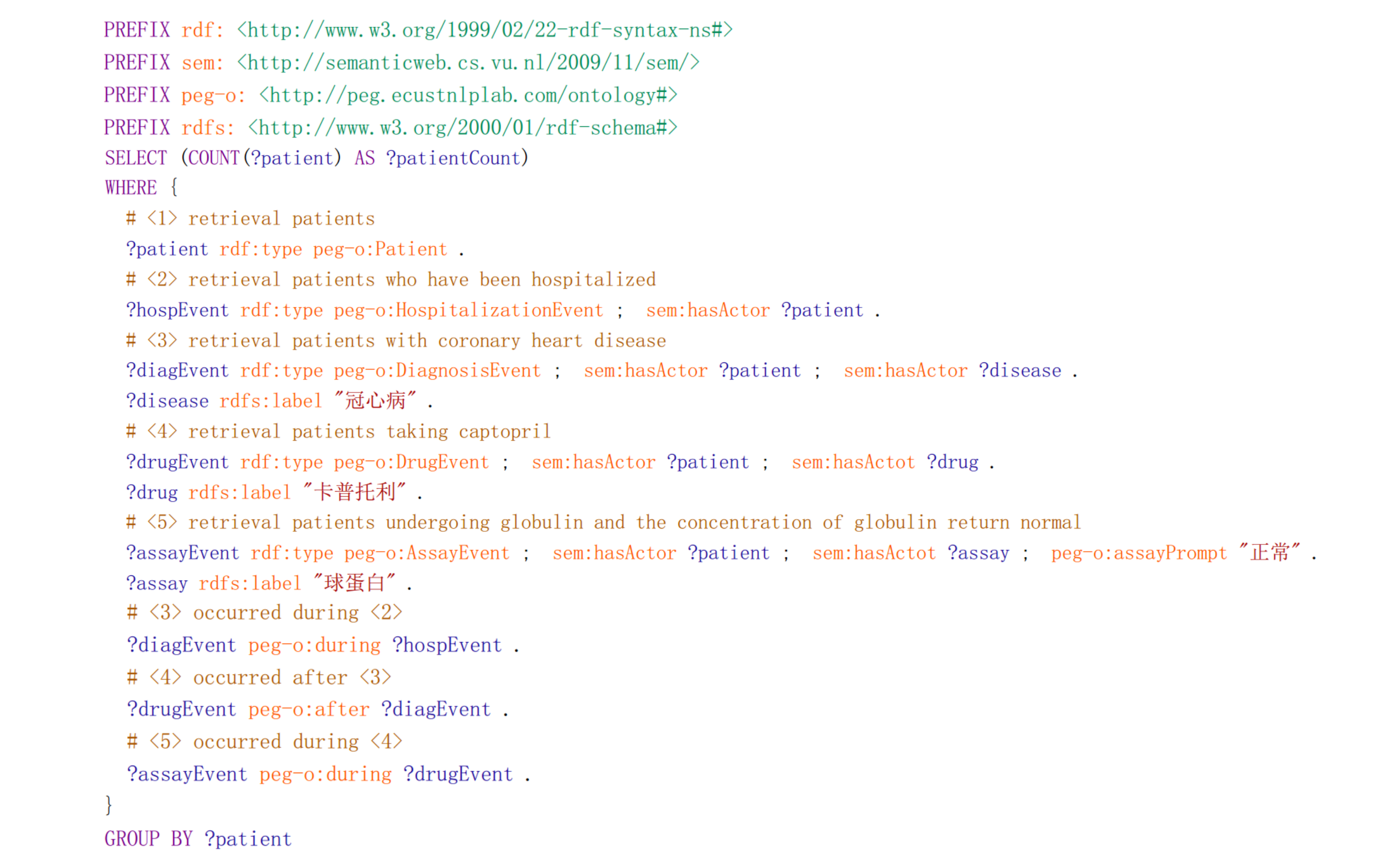}
\caption{An example of CME clinical problems upon PatientEG dataset. ``how many male(``男'') patients were  diagnosed with coronary heart disease (``冠心病'') during hospitalization and then took captopril (``卡普托利'') for treatment, during the medical treatment, their globulin (``球蛋白'') returned normal (``正常'')?''}\label{fig3_6}
\end{figure}
\end{CJK*}
Table.~\ref{tab2} lists part of clinical problems. Because of limited space, we only present a SPARQL query upon PatientEG dataset, shown as Fig.~\ref{fig3_6}, corresponding to the CME clinical problem in Table.~\ref{tab2}. And we put other SPARQL queries on our website.

\section{Related work}\label{relatedwork} 
Due to the practical significance, presentation of medical information has attracted considerable research effort, and a great number of solutions have been proposed in the literature. Kringelum et al. \cite{kringelum2016chemprot} built a dataset of chemical-protein-disease resources. Kuhn et al. \cite{kuhn2015sider} created a database of drugs and adverse drug reactions (ADRs). Schriml et al. \cite{schriml2011disease} presented a clinical coding system that classifies diseases according to certain characteristics of diseases and rules. Wang et al. \cite{wang2017pdd} linked disease and drug entities in MIMIC- III (Medical Information Mart for Intensive Care III) \cite{johnson2016mimic} to ICD-9 \cite{schriml2011disease} ontology and DrugBank \cite{wishart2017drugbank} to explore relations between entities such as drug-drug interactions. However, lacking of association with EMRs or omitting temporal information is far from adequate to fully unveil significance of EMRs. In general filed, Gottschalk and Demidova \cite{gottschalk2018eventkg} built a multilingual event knowledge graph from Wikidata \cite{erxleben2014introducing}, DBpedia \cite{lehmann2015dbpedia}, and YAGO \cite{mahdisoltani2013yago3}. To our best knowledge, none of the existing researches concentrate on applying event graph representation with temporal relations to EMRs in clinical field.

\section{Conclusion and Future Work}\label{conclusion}
In this paper, we presented the PatientEG model: a patient event graph representation to explicitly define medical entities, medical events and temporal relations for clinical research. Based on the proposed model, we describe a workflow of constructing a dataset from relational databases in hospital information systems. In order to normalize entity values and utilize synonymies, hyponymies and abbreviations of entities, we link them with a Chinese biomedical knowledge graph. Finally, we described the characteristics and example applications of PatientEG dataset.

Future work will focus on improving the coverage of instance matching via machine learning. Moreover, our intent is to populate our PatientEG model with more medical information in EMRs and expand our dataset so that clinical applications perform  better on it. Finally, we attempt to construct a natural language question answering system based on PatientEG dataset to address the gap between SPARQL query and natural language.

\section*{Acknowledgment}
This work was supported by the National Key R\&D Program of China for ``Precision Medical Research'' (No. 2018YFC0910500) and the National Natural Science Foundation of China (No. 61772201), and National Major Scientific and Technological Special Project for ``Significant New Drugs Development'' (No. 2018ZX09201008).
%
%
%
\bibliographystyle{splncs04}
\bibliography{bib.bib}
\end{document}